# HENO-MAC: Hybrid Energy Harvesting-based Energy Neutral Operation MAC Protocol for Delay-Sensitive IoT Applications

Sohail Sarang[1], Goran M Stojanović[1], Micheal Drieberg[2], Varun Jeoti[1] and Mikko Valkama[3]
[1]Faculty of Technical Sciences, University of Novi Sad, 21000 Novi Sad, Serbia
[2]Department of Electrical & Electronic Engineering, Universiti Teknologi PETRONAS, 32610 Seri Iskandar, Malaysia
[3]Department of Electrical Engineering, Tampere University, 33720 Tampere, Finland
[1]{sohail, goran, varunjeoti}@uns.ac.rs, [2]mdrieberg@utp.edu.my, [3]mikko.valkama@tuni.fi

*Abstract*— The Internet of Things (IoT) technology uses small and cost-effective sensors for various applications, such as Industrial IoT. However, these sensor nodes are powered by fixed-size batteries, which creates a trade-off between network performance and long-term sustainability. Moreover, some applications require the network to provide a certain level of service, such as a lower delay for critical data, while ensuring the operational reliability of sensor nodes. To address this energy challenge, external energy harvesting sources, such as solar and wind, offer promising and eco-friendly solutions. However, the available energy from a single energy source is insufficient to meet these requirements. This drives the utilization of a hybrid energy harvesting approach, such as the integration of solar and wind energy harvesters, to increase the amount of harvested energy. Nevertheless, to fully utilize the available energy, which is dynamic in nature, the sensor node must adapt its operation to ensure sustainable operation and enhanced network performance. Therefore, this paper proposes a hybrid energy harvesting-based energy neutral operation (ENO) medium access control (MAC) protocol, called HENO-MAC, that allows the receiver node to harvest energy from the solar-wind harvesters and adapt its duty cycle accordingly. The performance of the proposed HENO-MAC was evaluated using the latest realistic solar and wind data for two consecutive days in GreenCastalia. The simulation results demonstrate that the duty cycle mechanism of HENO-MAC effectively utilizes the harvested energy to achieve ENO and uses the available energy resources efficiently to reduce the packet delay for all packets and the highest priority packet by up to 28.5% and 27.3%, respectively, when compared with other existing MAC protocols.

## I. INTRODUCTION

The Internet of Things (IoT) is one of the most significant technologies that supports a wide range of applications, such as environmental and industrial process monitoring, healthcare, and agriculture [1]. Wireless sensor network (WSN) is an integral part of IoT and has gained significant attention due to its features such as small, inexpensive, and low-power nodes. These sensor nodes are powered by limited-size batteries [2]; as a result, they have a limited lifetime. In addition, batteries need to be changed regularly, which is time-consuming and hinders network operation.

Energy harvesting technology has enabled sensor nodes to collect external energy using different sources, such as solar, wind, radio frequency (RF), thermal, and mechanical, as shown in Figure 1. However, these ambient energy sources have different energy harvesting abilities that are dependent on various factors, such as time, weather conditions (sunny or rainy days), indoor or outdoor conditions, and atmospheric pressure [3]. Therefore, long-term network sustainability requires a balance between the available external energy and energy consumption, which is called energy neutral operation (ENO) [4]. When the external source does not provide sufficient energy under low harvesting conditions, the node relies on its stored energy in a battery, which can lead to a limited network lifetime, as shown in Figure 2. Therefore, the uncertainty in energy harvesting conditions and the amount of harvested energy from the single energy source is not adequate to achieve ENO, which can lead to limited network lifetime [1, 5]. This challenge motivates the integration of different energy harvesters to increase harvested energy to achieve long-term sustainability and desired performance goals [6].

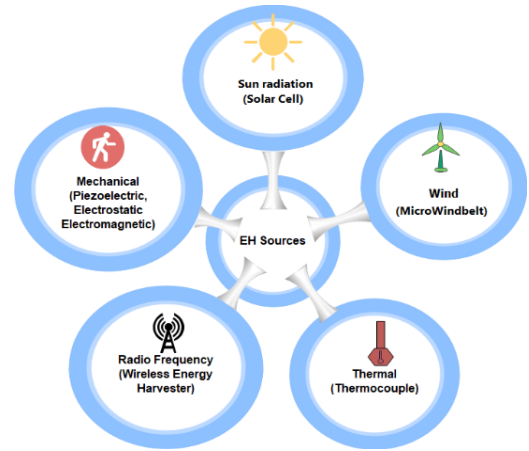

Fig. 1. Energy harvesting sources

In energy harvesting IoT, the key design objective of the medium access control (MAC) protocol is to use the harvested energy efficiently to maximize performance [7]. This can be achieved by developing smart energy allocation strategies based on the current energy harvesting conditions and overall available energy resources. In the literature, several energy-harvesting-aware MAC protocols have been developed for different applications, such as environmental monitoring, agriculture, and delay-sensitive applications [8-10]. They incorporated harvested energy information to set different parameters, such as access probability, duty cycle, node priority, and relay node selection. However, the performance of existing protocols primarily relies on a single harvesting source; thus, they may not provide sufficient energy at all times. Moreover, in dynamic harvesting scenarios, such as rainy days or low wind conditions, sensor nodes may be required to compromise network performance to sustain their operation in the absence of external energy. This motivates the exploitation of the hybrid energy harvesting approach to increase the energy resources of sensor nodes, which can enhance the overall reliability and sustainability of the network. Furthermore, the hybrid energy harvester can provide a sufficient amount of harvested energy over different periods, which can help achieve ENO. Thus, the sensor node can utilize the ENO and available energy more effectively to enhance the performance. For this purpose, the solar-wind hybrid energy harvester is a good choice for powering nodes in IoT applications [12]. This work introduces a solar-wind hybrid energy source that allows sensor nodes to continuously replenish energy to ensure long-term sustainability. Furthermore, it presents an ENO-based duty cycle mechanism to efficiently utilize hybrid harvested energy to optimize network performance.

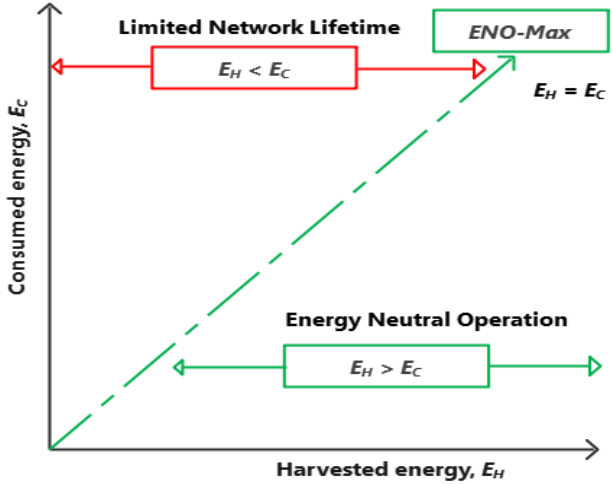

Fig. 2. ENO condition [11]

The main contributions of the work are:

- This work uses a solar-wind hybrid harvester that enables sensor nodes to harvest energy continuously from the surrounding environment.
- An ENO-based duty cycle mechanism is introduced that adjusts the duty cycle based on the ENO and available energy resources to enhance performance.
- The performance of the proposed HENO-MAC has been evaluated in GreenCastalia [13] using the latest actual solar irradiance and wind speed data [14].
- Its performance comparison with three EH MAC protocols, namely ENCOD-MAC [15], QPPD-MAC [10], and QAEE-MAC [16] in terms of average delay for all packets and priority packets.

The structure of this paper is as follows: Section II reviews the related works. Section III outlines the proposed HENO-MAC protocol. Section IV presents the performance evaluation and comparison. Finally, Section V provides the conclusions and future work.

## II. RELATED WORKS

Currently, there is an increasing trend in the design of energy-harvesting-aware protocols and algorithms. The key objective is to schedule tasks based on harvested energy to maximize performance. Focusing on MAC protocols, for example, EEM-MAC [17] considers the individual duty cycle of a node and regulates its listening time based on the remaining energy. Similarly, MTE-EHWSN [18] introduced a multi-threshold approach that assigns different duty cycle values based on the residual energy. RF-AASP-MAC [19] incorporates traffic and available RF energy conditions to regulate the wake-up schedule of the node. In OD-MAC [11], the node considers the harvesting rate to adapt to the beacon interval and sensing time to improve performance. Similarly, in LEB-MAC [20], the node sets its next wake-up schedule according to the harvesting conditions to achieve load balancing among nodes. PSL-MAC [8] utilizes energy-harvesting information to prioritize packet transmission. AE-MAC [21] sets different contending priorities based on the energy levels of the nodes. MEH-MAC [22] supports node mobility in EH-WSNs. MDP-SHE-WSNS [23] determines the packet transmission based on the energy level. QAEE-MAC [16] supports the priority of data packets and considers the energy state during data transmission. In EH-TDMA-MAC [24], the authors evaluated network throughput using different harvesting rates. The receiver in QPPD-MAC [10] adjusts its duty-cycle-based available energy to reduce packet delay under high-harvesting conditions. ENCOD-MAC [15] helps the receiver to operate close to the ENO and sets its duty cycle according to available energy. ED-CR [25], HM-RIMAC [9], R-MAC [26], and PBP-MAC [27] account for the residual energy in adjusting the listening time of the node and select the cluster and relay nodes accordingly. RF-AASP-MAC [19] uses traffic conditions and energy information to adjust the node operation. However, to the best of our knowledge, the performance of existing MAC protocols has not yet incorporated and been tested using hybrid energy sources. As a result, they may be unable to effectively harness a significant amount of harvested energy when employed with a hybrid energy harvester. Therefore, there is a need to develop a new MAC protocol for energy harvesting IoT that can harness the combined energy from the hybrid energy source and make use of available energy resources efficiently to achieve long-term network sustainability and improved performance.

## III. PROPOSED HENO-MAC PROTOCOL

This section explains the HENO-MAC protocol and its features. First, basic communication is provided, which describes the communication between the receiver and two sender nodes. Then, an ENO-based duty cycle mechanism is described in which the receiver adjusts its duty cycle according to the harvested energy using a solar-wind hybrid source.

### A. Basic communication overview of HENO-MAC

The basic communication overview between the receiver and two sender nodes is shown in Figure. 3. We consider a case in which each sender node can assign four different priority levels, $P_4$, $P_3$, $P_2$, and $P_1$, to the data packet. $P_4$ defines time-critical data with an acyclic nature, such as an emergency alarm, which is always critical and needs to be forwarded faster compared to other priority packets. Other priority levels, $P_3$, $P_2$, and $P_1$, refer to real-time (most important), on-demand (important), and periodic monitoring, respectively. The delay requirement for each priority level depends on the application. Initially, the receiver node broadcasts a wake-up beacon (WB), which contains two fields: self-address (SA) and energy state ($E_s$), as shown in Figure 4(a). Then, it starts a waiting timer ($T_w$) to accept the incoming Tx–beacon (TxB) packets from both sender nodes. However, after receiving the WB, the sender nodes check the $E_s$ field to determine whether the receiver has sufficient energy. If yes, it starts channel sensing immediately. If the channel is free, the node sends its TxB, which includes the packet priority (P) and destination address (DA) fields, as shown in Figure 4(b). Through TxB, the sender node informs the receiver of the priority of its data packet and transmits it. The receiver adjusts the timer $T_w$ value according to the priority set in TxB. It cancels timer $T_w$ if the highest priority packet, that is, $P_4$, arrives. This is because the receiver has already received a TxB request from the sender that contains the highest $P_4$ priority level; hence, there is no need to wait for other TxBs. Subsequently, the receiver broadcasts RxB, which indicates the address of the selected sender (SS). Figure 4(c) shows the frame structure of RxB. Then, the selected sender node sends the data to the receiver and waits for acknowledgment (ACK), whereas the other nodes go to sleep to conserve energy.

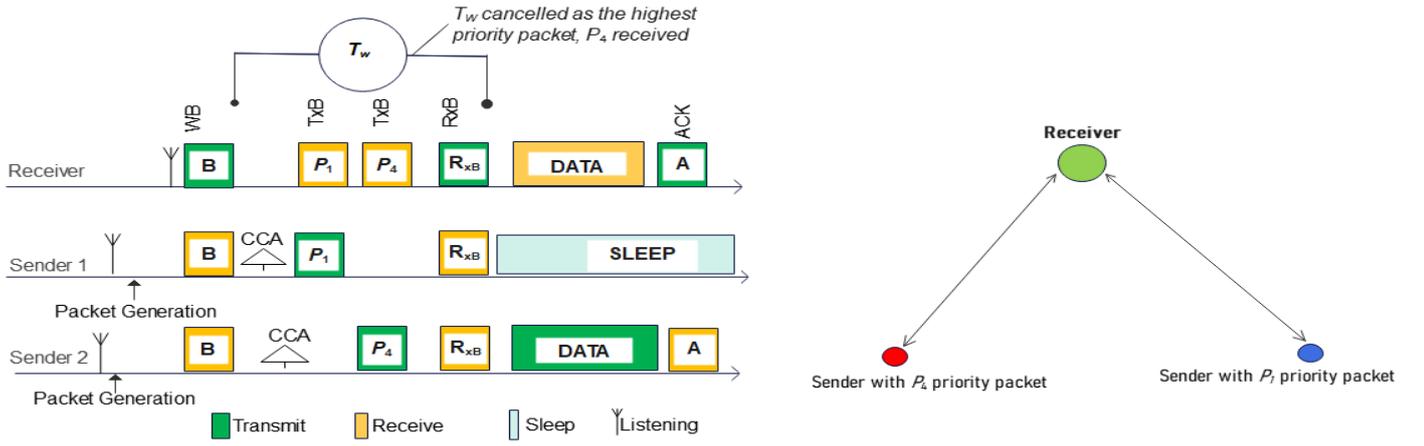

Fig. 3. Communication overview of HENO-MAC

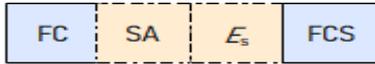

(a) Wake-up Beacon (WB)

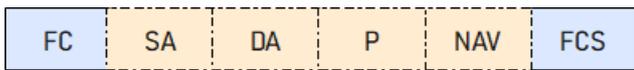

(b) Tx Beacon (TxB)

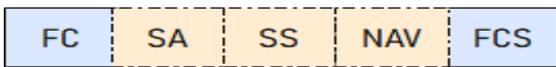

(c) Rx Beacon (RxB)

Fig.4 Frame structure of different beacons used in establishing communication between receiver and sender nodes. FC and FCS correspond to the Frame Control and Frame Check Sequence fields, respectively, which are taken from IEEE 802.15.4 standards

### B. Proposed Duty Cycle Mechanism

A hybrid energy harvester enables a node to harvest a sufficient amount of energy from the sensor node, leading to long-term network sustainability. Moreover, to harness the combined energy and use it effectively, the proposed HENO-MAC introduces an ENO-based duty cycle, $d_c$ mechanism which adjusts the receiver duty cycle according to ENO and available energy. The $d_c$ value is set to 1 when ENO is achieved. This helps the receiver remain active for a longer period to accept incoming packets, and as a result, packet delay is reduced in the network. The hybrid energy harvester allows the node to collect energy continuously at various harvesting rates. This drives the use of available energy resources such as sufficient remaining energy ($RE_{Total}$) effectively to maximize the performance. Thus, when $RE_{Total}$ is between 50-100%, the duty cycle $d_c$ is set to 1 which further helps the receiver to remain active over different periods without achieving ENO. If the receiver does not achieve ENO and its $RE_{current}$ ranges between 10-50%. Then it uses Eq. (6) to calculate $d_c$. In the worst harvesting scenario, when $RE_{Total}$ falls below the threshold energy level ($E_{th}$) i.e., 10%, the node is set to operate on a very low $d_c$ of 0.05 to avoid complete energy depletion.

The $RE_{Total}$ can be computed as follows:

$$RE_{total} = (RE_{current} + E_h + E_w - E_T) \quad (1)$$

where $RE_{current}$, $E_T$, $E_h$, and $E_w$ represent current remaining energy, total consumed energy, and harvested energy from solar and wind in joule (J), respectively. The average solar energy $E_h$ per slot is computed as [28]:

$$E_h = A \times \eta \times I_s \times d_s \quad (2)$$

where $A$, $\eta$, $I_s$ and $d_s$ denote solar panel area in m², panel efficiency, solar irradiance (W/m²), and duration of a timeslot in seconds. The average harvested energy from the wind micro turbine, $E_w$ is calculated as follows:

$$E_w = 0.5 \times v^3 \times A_w \times \rho \times C_p \times d_s \quad (3)$$

where $v$, $A_w$, $\rho$ and $C_p$ represent the wind speed in m/s, rotor swept area in m², air density in kg/m³ and power coefficient, respectively.

$E_T$ of a node can be computed as follows

$$E_T = \sum_{i=0}^{n} P_i \times t_i \quad (4)$$

where $p$, $t$, and $n$ denote the power consumption rate of the radio in state $i$, time spent in state $i$, and the number of states, respectively. Based on the state of the radio, and the time duration that the node spent in a specific radio state: transmission, reception, idle listening, and sleeping energy is drawn from the battery.

$RE_{total}$ in percentage is given as follows:

$$RE_{total} (\%) = \frac{RE_{total}}{E_{max}} \quad (5)$$

where $E_{max}$ denotes the maximum capacity of the battery. Moreover, Table 1 and Eq (6) illustrate the $d_c$ values corresponding to ENO condition and $RE_{total}$ level, where $E_c$ represents the energy consumption of a node when it operates continuously at the maximum duty cycle of 1 for an hour i.e., 224 J. The node achieves an ENO condition when the combined value of $E_h$ and $E_w$ is more than the required energy, $E_c$.

$$d_c = \frac{RE_{total} - E_{th}}{100\% - E_{th}} \quad (6)$$

The obtained $d_c$ value is utilized to compute the sleep time ($T_{sleep}$) as given below

$$T_{sleep} = \frac{T_{listen} \times (1 - d_c)}{d_c} \quad (7)$$

where $T_{listen}$ represents the listening time.

TABLE I. DUTY CYCLE ADJUSTMENT

| Parameter | Value |
|---|---|
| ENO: $(E_h + E_w) \geq E_c$ | 1 |
| $50 \leq RE_{total} \leq 100\%$ | 1 |
| $10 \leq RE_{total} < 50\%$ | Using Eq. (6) |
| $0 < RE_{total} < 10\%$ | 0.05 |

## IV. PERFORMANCE EVALUATION AND COMPARISON

This section evaluates the performance of the proposed HENO-MAC using GreenCastalia 3.3 [13] under the latest 13th and 14th June 2022 solar irradiance and wind speed data obtained from NREL [14]. The performance has been evaluated in terms of average delay for all packets and for the highest priority packets. The star network topology is implemented, as given in Figure 5. The receiver is positioned at the center, whereas the seven sender nodes are randomly distributed within a 30 m × 30 m area. Moreover, the receiver node is equipped with a 3000 mAh rechargeable battery, a solar cell of size 7.7 cm², and a small wind turbine with a rotor diameter of 5 cm. The sensor and radio parameters are adopted from TelosB and CC2420 transceiver. Each sender node generates one packet per second, assigns a priority based on a random number generated between 0 and 1, and follows a *p*-persistent CSMA scheme to forward data packets to the receiver. The simulation results have been compared with those of three other MAC protocols, namely ENCOD-MAC, QPPD-MAC, and QAEE-MAC. Table II shows the simulation parameters.

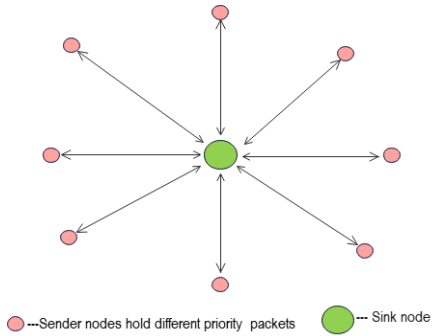

Fig. 5. Network topology

TABLE II. SIMULATION PARAMETERS

| Parameter | Value |
|---|---|
| Simulation time | 172800 s |
| $A$ | 7.7 cm² |
| $\eta$ | 22% |
| $I_s$ | Variable |
| $v$ | Variable |
| $A_w$ | 5 cm |
| $C_p$ | 0.1 |
| $\rho$ | 1.25 kg/m³ |
| No. of timeslots in a day | 24 |
| $d_s$ | 3600 s |
| $E_{max}$ | 3000 mAh |
| Initial battery level | 25% |
| Sender nodes | 1 to 7 |
| Coverage area | 30 m × 30 m |
| TelosB operating voltage | 2.1 V |
| Data rate | 250 kbps |
| Packet size | 28 bytes |
| TxB size | 14 bytes |
| RxB size | 13 bytes |
| ACK size | 11 bytes |
| WB size | 9 bytes |
| Slot time | 0.32 ms |
| $T_w$ | 5 ms |

Fig. 6 shows the average combined and separately harvested energy from the solar and wind sources. The energy from solar depends on solar irradiation; hence, it is only available during the daytime. Similarly, the wind energy changes significantly owing to the turbulence and speed variations. Consequently, the output energy fluctuates rapidly and significantly over different periods of the day. It can be seen that solar energy gradually rises after sunrise at 6 AM and reaches a maximum value of 585 J at approximately 2 PM on the first day. It then starts declining and ultimately reaches zero at sunset at approximately 9 PM. On the same day, the wind source provided a maximum energy of 225 J around 6 AM and then exhibited rapid fluctuations throughout the day. The harvested energy from the solar and wind sources shows a similar trend on the following day. It can also be noticed that the combined energy from both sources reached the maximum of 724 J and 657 J on the first and second days, respectively.

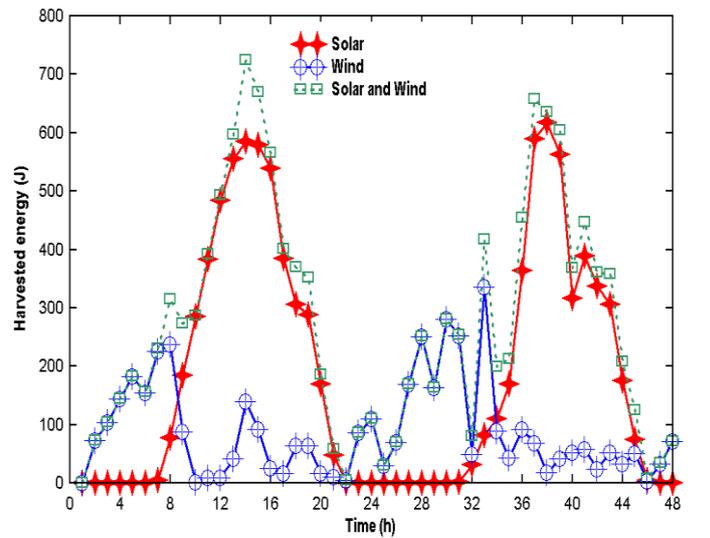

Fig. 6. Average harvested energy from solar and wind sources based on NREL data for two consecutive days (13-14 June 2022).

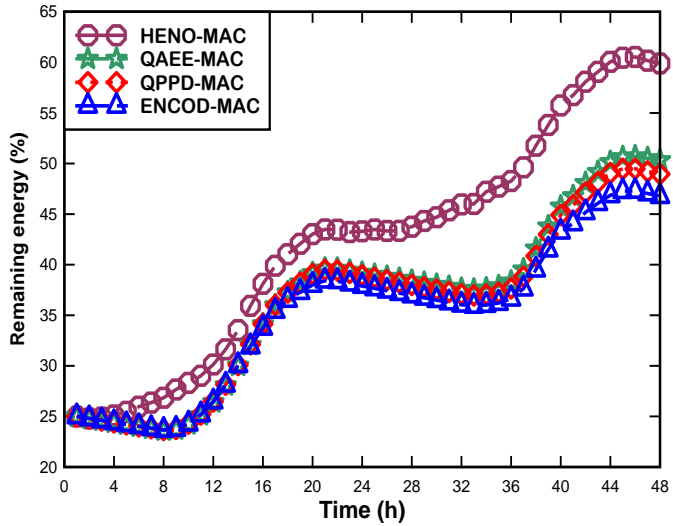

Fig. 7. Receiver remaining energy, $RE_{total}$ level

The total remaining energy, $RE_{total}$ is computed based on harvested energy, and energy consumption of the node, as given in Eq. (1). The initial battery level is set to 25% for all protocols, as shown in Fig. 7. In HENO-MAC, the node starts harvesting energy immediately from the wind source and as a result, $RE_{total}$ increases gradually. On the other hand, the receiver in other protocols utilizes a single harvesting source, that is, solar, and as a result, the harvested energy is not available until sunrise. On the first day, $RE_{total}$ increases to a maximum value of 43.5%, 39.6%, 39.5%, and 37.8% in HENO-MAC, QAEE-MAC, QPPD-MAC and ENCOD-MAC, respectively. After that, a similar trend can be seen on the second day where $RE_{total}$ reaches 59.8%, 50%, 48.7%, and 50% until the end of the second day, respectively.

Fig. 8 represents the duty cycle, $d_c$ in all protocols. It can be seen that the proposed HENO-MAC, QPPD-MAC, and ENCOD-MAC adjust their receiver $d_c$ according to the energy harvesting conditions. However, the QAEE-MAC operates on a fixed $d_c$. Moreover, HENO-MAC combines the harvested energy from the solar-wind harvester and adjusts $d_c$ more aggressively when sufficient energy resources are available. This is because the proposed $d_c$ mechanism enables the receiver to increase its $d_c$ to the maximum, that is, 1, when it achieves the ENO condition. It is noted that the hybrid energy approach enables the receiver to remain in the ENO state for a total of 24 h over different periods. This helps the receiver to remain active most of the time, which helps reduce the delay for incoming packets. Furthermore, a few short-term fluctuations in $d_c$ is because of sudden changes in available energy over different periods due to the surrounding conditions. The receiver in HENO-MAC promptly adapts to quick fluctuations in the harvested energy by adjusting its $d_c$ value accordingly. Moreover, when the average combined energy in a slot is insufficient, the receiver sets its $d_c$ based on the stored energy. Thus, if the available energy level is equal to or greater than 50%, the duty cycle is increased to 1 to further reduce the delay.

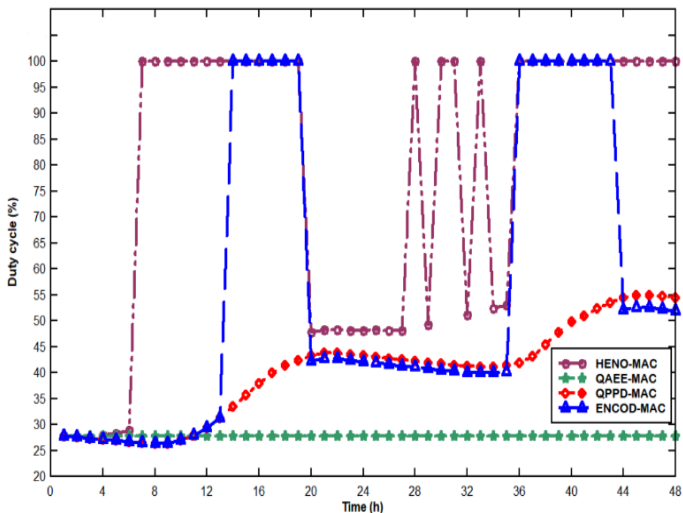

Fig. 8. Receiver $d_c$ according to remaining energy, $RE_{total}$

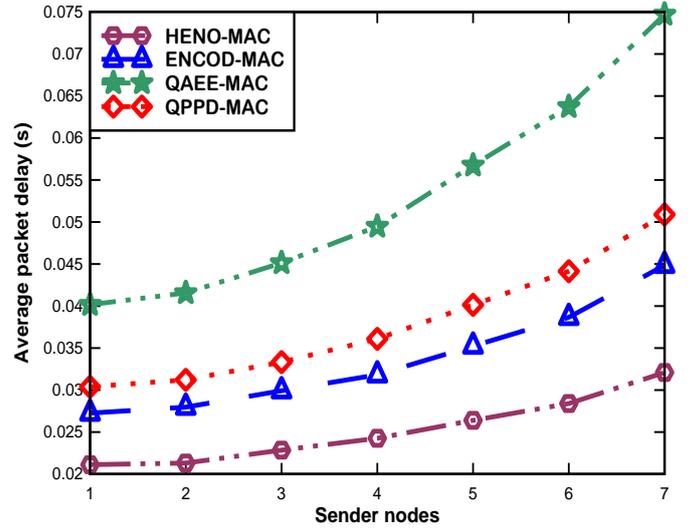

Fig. 9. Average delay of all packets

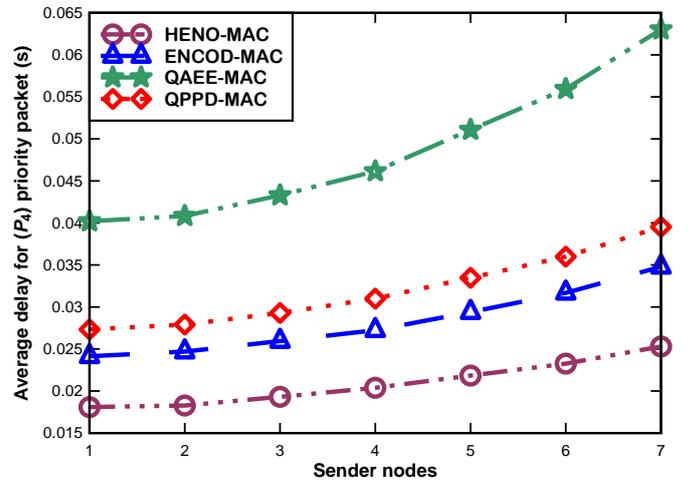

Fig. 10. Average delay for the highest, $P_4$ priority packets

The packet delay represents the time duration from the generation of the data packet to its arrival at the receiver. Figures 9 and 10 show the average packet delay for all data packets and the highest $P_4$ priority packet, respectively. The results show that the proposed HENO-MAC protocol provides the lowest average delay of all packets and $P_4$ priority packets and improvements of up to 28.5% and 27.3%, respectively, when compared to the other MAC protocols. This is because the receiver in HENO-MAC adapts its duty cycle according to the ENO and the remaining energy level. Consequently, a sufficient amount of harvested energy allows the receiver to remain active for a longer time for incoming data packets, resulting in significantly lower delays. However, in other MAC protocols, the receiver relies on a single energy source and thus does not have sufficient energy resources or a mechanism to optimize performance.

## V. CONCLUSION AND FUTURE WORK

Energy harvesting technology has gained significant attention in IoT applications. However, the uncertainty in harvesting energy motivates the design of energy harvesting aware protocols. This work presented an ENO-based MAC protocol that enables the sensor node to collect energy from a hybrid solar-wind harvester to achieve the ENO in the network. Furthermore, it allows the receiver to set its duty cycle according to the ENO and the available energy resources. In a high harvesting scenario, the receiver harvests a significant amount of energy, and as a result, it aggressively increases the duty cycle by reducing the sleep time to improve performance. In low energy conditions, it accounts for the remaining energy to set the duty cycle accordingly to ensure sustainable operation. The performance of HENO-MAC is tested in GreenCastalia under realistic harvesting conditions for two consecutive days. The results demonstrate that the HENO-MAC is able to incorporate the combined harvested energy effectively in making energy harvesting aware duty cycling, to reduce the packet delay significantly. In summary, the proposed HENCO-MAC effectively contributes to the core objective of energy harvesting IoT to achieve long-term sustainability and effective use of harvested energy to improve application performance, especially for delay-sensitive applications. The future work will include the extension and testing of HENO-MAC for multi-hop scenarios under realistic conditions.

ACKNOWLEDGMENT

This work was supported by the European Union's Horizon 2020 Research and Innovation Programme under the grant agreement No 854194.